\documentclass[aip,reprint]{revtex4-1}
\usepackage{graphicx}
\usepackage[T1]{fontenc}
\usepackage{bm}
\usepackage{amsmath}

\draft 

\begin{document}

\title{An exhaustive list of isotropic apocalyptic scenarios} 

\author{S.L.Parnovsky}
\email[]{par@observ.univ.kiev.ua}
\affiliation{Astronomical Observatory, Taras Shevchenko National University of Kyiv, 4 
Observatorna Str., 04053, Kyiv, Ukraine}

\date{\today}

\begin{abstract}
We study the possible types of future singularities in the isotropic homogeneous
cosmological models for the arbitrary equation of state of the contents of the Universe. 
We obtain all known types of these singularities as well as two new types using a
simple approach. No additional singularity types are possible. We name the new 
singularities type ``Big Squeeze'' and ``Little Freeze''.
The ``Big Squeeze'' is possible only in the flat Universe after a finite time 
interval. The density of the  matter and dark energy tends to zero and its 
pressure to minus infinity. This requires the dark energy with a specific
equation of state that has the same asymptotical behaviour at low densities 
as the generalised Chaplygin gas. The ``Little Freeze'' involves an eternal 
expansion of the Universe. Some solutions can mimic the $\Lambda$CDM model.
\end{abstract}


\maketitle 

\section{Introduction}

During almost a century, cosmologists considered only two possible scenarios of the future
of our Universe -- an eternal expansion of open or flat Universe or future recollapse of the closed Universe
with the ``Big Crunch''. Nowadays we know that the Universe contains
not only several types of matter, including the dark matter, baryonic matter and massless particles,
but also the mysterious dark energy (DE). We know about its existence only for the last few decades.
Honestly, we know very little about DE properties, in particular about the DE equation of state. 

Even for the simplest type of the DE equation of state 
\begin{equation}\label{e0}
p=w\rho
\end{equation}
with $w=const$, where $p$ is the pressure
and $\rho$ is the mass density, the Universe can meet its end in absolutely different way. If
$w<-1$ we deal with so-called phantom energy. In this case during the finite time period the matter and energy density,
the Hubble parameter $H$ and the scale factor of the Universe $a$ increase to infinity. Such type of possible
future singularity was discovered in [\onlinecite{r1}] and called ``Big Rip''. 

Note that the latest
estimations of the $w$ value do not reject this possibility. The data on
the cosmic microwave background spectra from the Planck and WMAP satellites together with ground measurements and data 
from baryonic acoustic oscillations (BAO) provide the estimation $w=-1.13^{+0.23}_{-0.25}$ at 95\% confidence 
level (CL) [\onlinecite{r2}].
The 9-year data from the WMAP satellite plus the determination of the Hubble constant and BAO data provide 
estimations $w=-1.073^{+0.090}_{-0.089}$ for the flat Universe and $w=-1.19\pm 0.12$ for 
the non-flat Universe [\onlinecite{r3}] at 68\% CL. Adding 472 type Ia supernovae data improves these estimations to
$w=-1.084\pm 0.063$ and $w=-1.122^{+0.068}_{-0.067}$, respectively [\onlinecite{r3}].

Thus, the possibility of the ``Big Rip'' sealing the fate of the Universe is not to be taken lightly.
This is not the only theoretically possible type of cosmological singularity except ``Big Bang'' and ``Big Crunch''. 
Their first classification
was carried out in the paper [\onlinecite{r4}]. Four possible types were found for the singularities at $t=t_0$ 
with finite $t_0$. They include:
\begin{description}
\item[$\bullet$ Type I] $a,\rho,|p|\to \infty$ (``Big Rip'')
\item[$\bullet$ Type II] $a\to a_0;\rho \to \rho_0;|p|\to \infty$ (``sudden'')
\item[$\bullet$ Type III] $a\to a_0;\rho,|p|\to \infty$ (it was named ``Big Freeze'' in [\onlinecite{r5}])
\item[$\bullet$ Type IV] $a\to a_0;\rho,|p|\to 0$ and higher derivatives of the Hubble parameter $H$ diverge.
\end{description}
In the more recent classification [\onlinecite{r6}], type IV singularities are divided into type IV and 
type V, introduced in [\onlinecite{r12}], but we stick to the classification [\onlinecite{r4}]. 
There are some singularities with $t_0=\infty$, too. 
The ``Little Rip'' singularity [\onlinecite{r7}], similar to the ``Big Rip'',
but with eternal expansion is among them. 

Some types of singularities were found and demonstrated for some specific equations of state. Cosmologists considered
the particular cases of the phantom 
generalised Chaplygin gas equation of state in [\onlinecite{r5}], tachyon field in [\onlinecite{r8}], scalar fields with specific 
potentials, etc. Naturally, a question arose, whether all the possible singularity types have been considered. 

In this article we 
try to give an exhaustive answer to this question for the isotropic and homogeneous Universe. To make it worse, 
in addition to unknown DE equation of state we have three possible signs of space curvature. We are interesting in the complete
list of the possible types of future singularities for an arbitrary equation of state for three signs of space curvature. The possible
singularity types for the flat Universe were considered in the paper [\onlinecite{r6}], but we find a new one.
We consider an arbitrary equation of state $p(\rho)$ without any constrains except 
$\rho\ge 0$. In particular we do not use the strong energy condition $\rho+3p>0$.

\section{The search for future singularities in FLRW Universe}
We consider the 
homogeneous isotropic Universe with the Friedmann-Lema\^{i}tre-Robertson-Walker (FLRW) metric
\begin{equation}\label{e1}
ds^2=dt^2-a(t)^2\left[ d\chi^2+F^2(\chi)dO^2\right],
\end{equation}
where $a(t)$ is the scale factor, $dO^2=d\Theta^2+\cos^2(\Theta)d\varphi^2$ is the distance element on a unit sphere, 
$F(\chi) = \sin(\chi)$ and $k=1$
for the closed Universe, $F(\chi) = \sinh(\chi)$ and $k=-1$ for the open one, and $F(\chi) = \chi$ and $k=0$ for the spatially flat models. 
We use the system of units in which $G = 1$ and $c = 1$. This Universe is 
filled by all kinds of matter and dark energy with a mass density $\rho$ and an effective pressure $p(\rho)$. In this system of units 
the energy density $\varepsilon$ coincides with $\rho$. The Einstein equations for the metric (\ref{e1}) reduce to the 
well-known Friedmann equations. We need the expression for the Hubble parameter $H=a^{-1}da/dt$
\begin{equation}\label{e2}
H^2=\frac{8\pi}{3}\rho-\frac{k}{a^2}
\end{equation}
and the hydrodynamical equation or the energy conservation equation
\begin{equation}\label{e3}
\frac{d\rho}{dt}=-3(\rho+p)H.
\end{equation}

The Friedmann equation for the scale factor 
\begin{equation}\label{e4}
\frac{d^2 a}{dt^2}=-\frac{4\pi}{3} a(\rho + 3p)
\end{equation}
follows from the equations (\ref{e2}) and (\ref{e3}). 

\subsection{Flat model} 
We start from the flat model with $k=0$. The equation (\ref{e2}) provides the expression $H=(8\pi \rho /3)^{1/2}$. After substituting 
it into (\ref{e3}) we obtain a simple equation with the solution
\begin{equation}\label{e5}
\Delta t=t_0-t_1=-\frac{1}{2(6\pi)^{1/2}} \int_{\rho_1}^{\rho_0} \frac{d\rho}{\rho^{1/2}\left(\rho+p(\rho)\right)}.
\end{equation}
Here the subscript 1 corresponds to the initial parameters (i.e. $t_1$ is ``now'') and the subscript 0 corresponds to the parameters 
of the Universe in the future at time $t_0$ after a time interval $\Delta t$. We will denote the instant of time of any terminal cosmological 
singularity as $t_0$, and use (\ref{e5}) to analyse their properties. After finding the dependence $\Delta t(\rho)$ we find the 
inverse function $\rho(\Delta t)$ and $H(\Delta t)$, the integration of the last one gives $\ln(a)$.

The first thing to check is the finiteness of $\Delta t$. If the integral 
in (\ref{e5}) diverges we obtain $t_0=\infty$ and this case deals with the asymptotic evolution in the future. An example of such 
solution is the ``Little Rip'' [\onlinecite{r7}]. 

We are going to go over all possible types of singularity. We consider three possible cases for $\rho_0$. It can be infinite, finite 
and nonzero, or equal to zero. Let us consider it one by one.

\subsubsection{Infinite terminal density}

Let us start with a well-known ``Big Rip'' singularity to illustrate our approach. 
We consider the equation of state (\ref{e0}). If $w=-1$ we deal with the 
effective cosmological constant. According to (\ref{e3}) in this case the density and the pressure are constant. If $w>-1$ the values of $\rho$ and $H$ decrease 
in time because of (\ref{e3}). If $w<-1$ the values of $\rho$ and $H$ increase due to (\ref{e3}) and become infinite 
at time $t_0$. Equation (\ref{e5}) gives us in this case the relations
\begin{equation}\label{e6}
\rho_1=\frac{1}{6\pi (1+w)^2 \Delta t^2},\; H=\frac{2}{3|1+w|\Delta t}.
\end{equation}
This is the so-called ``Big Rip'' case [\onlinecite{r1}]. The scale factor of the Universe diverges $a\propto \Delta t ^{-\frac{2}{3|1+w|}}$.

A somewhat similar case is when $w$ is not constant, but asymptotically tends to $-1$: $\rho/p\xrightarrow [\rho \to \infty]{}-1$.
Let us assume that it follows the power law
\begin{equation}\label{e7}
\rho+p\xrightarrow [\rho \to \infty]{}-A\rho^\alpha
\end{equation}
with $\alpha < 1$, $A=const$. The integral in (\ref{e5})
is finite at $1/2<\alpha<1$. In this case we have the ``Big Rip'' with $H\propto \Delta t^{1/(1-2\alpha)}$, 
$\ln a\propto \Delta t^{2(1-\alpha)/(1-2\alpha)}$. It occurs later and has a sharper shape for the same initial value $\rho_1$ in 
comparison with the equation of state (\ref{e0}). 

If $\alpha<1/2$ the integral in (\ref{e5}) becomes divergent and we have to put 
$t_0=\infty$. This is the so-called ``Little Rip'' introduced in [\onlinecite{r7}]. In this case we rewrite (\ref{e5}) in the 
form
\begin{equation}\label{e8}
\Delta t=t-t_1=-\frac{1}{2(6\pi)^{1/2}} \int_{\rho_1}^{\rho(t)} \frac{d\rho}{\rho^{1/2}\left(\rho+p(\rho)\right)}.
\end{equation}
This case corresponds to an eternally accelerating expansion of the Universe: $H\propto t^{1/(1-2\alpha)}$, 
$\ln a\propto t^{2(1-\alpha)/(1-2\alpha)}$.

In the intermediate case $\alpha=1/2$ we must take into account a possible logarithmic divergence and consider the equation of state
with the asymptote $\rho+p\xrightarrow [\rho \to \infty]{}-A\rho^{1/2}(\ln \rho)^\beta$. At $\beta>1$ we deal with the unconventional 
``Big Rip'' with $\ln\rho \propto \Delta t^{1/(1-\beta)}$, at $\beta<1$ we deal with the ``Little Rip'' with $\ln\rho \propto t^{1/(1-\beta)}$.
At $\beta=1$ we consider the equation of state with the asymptotic $\rho+p\xrightarrow [\rho \to \infty]{}-A\rho^{1/2}(\ln \rho)
(\ln \ln \rho)^\gamma$. There is the ``Big Rip'' with $\ln\ln\rho \propto \Delta t^{1/(1-\gamma)}$ at $\gamma<1$ and the ``Little Rip'' with 
$\ln\ln\rho \propto t^{1/(1-\gamma)}$ at $\gamma>1$. If $\gamma=1$ we can go on with this way and consider the asymptotic 
$\rho+p\xrightarrow [\rho \to \infty]{}-A\rho^{1/2}(\ln \rho) (\ln \ln \rho) (\ln \ln \ln \rho)^\delta$, etc. Similar results were 
obtained in [\onlinecite{r7}].

So far we considered cases with $a\xrightarrow [\rho \to \infty]{}\infty$, but this is not requared. For example, a type III 
singularity, which was named ``Big Freeze'' in the paper [\onlinecite{r5}], has finite $t_0$ and $a_0$ values, but 
$\rho,H,|p|\xrightarrow [t \to t_0]{}\infty$. Let us consider this type of singularity. From $H=a^{-1}da/dt\xrightarrow [t \to t_0]{}\infty$
and $a(t)\xrightarrow [t \to t_0]{}a_0$ we see that $a(t)$ is regular, but $da/dt$ diverges at $t=t_0$. This is possible if the scale 
factor has a power-law asymptote 
\begin{equation}\label{e9}
a(t)\xrightarrow [t \to t_0]{}a_0-B(t_0-t)^\lambda
\end{equation}
with $0<\lambda<1$. This yilds $H\xrightarrow [t \to t_0]{}\lambda B(t_0-t)^{\lambda-1}/a_0$. From (\ref{e2}) we obtain for this case 
$\rho(t) \propto (t_0-t)^{2(\lambda-1)}$. After substituting these expressions in (\ref{e3}) we get 
$\rho(t)+p(t) \propto (t_0-t)^{\lambda-2}$. This corresponds to the equation of state (\ref{e7}) with $\alpha=\frac{2-\lambda}{2-2\lambda}$, 
$\lambda=\frac{2\alpha-2}{2\alpha-1}$. In this case $\alpha>1$ and $|p| \propto \rho^\alpha\gg \rho$ in the vicinity of the singularity.

Let us consider this type of singularity directly from (\ref{e5}). If $\rho \xrightarrow [t \to t_0]{}\infty$ but $\rho/p\xrightarrow [\rho \to \infty]{}0$, e.g. 
$p(\rho) \xrightarrow [\rho \to \infty]{} -A\rho^\alpha$ with $\alpha > 1$, $A=const$ we also have a singularity with 
$H\propto \Delta t^{1/(1-2\alpha)}$, $\ln a\propto \Delta t^{2(1-\alpha)/(1-2\alpha)}=\Delta t^\lambda$. Note that at $\alpha < 1$ 
we get the ``Big Rip'' case considered above. But in the case of the ``Big Freeze'' singularity the scale factor tends to some constant
value. Thus, we can study the ``Big Freeze'' either starting from the asymptotic behaviour of the equation of state (\ref{e7}) with $\alpha>1$
or from the asymptote (\ref{e9}). The asymptotic behaviour of the parameters of the 
Universe $p(t)(t-t_0)\rho(t)^{-1} \xrightarrow [t \to t_0]{} const$ near the type III singularity follows from the above-mentioned 
asymptotes.

If we deal with the power law (\ref{e9}) for the scale factor with some noninteger $\lambda>1$ we have no ``Big Freeze'' 
singularity, but some higher derivatives of $H$ diverge. If $1<\lambda<2$ both parts of the Friedmann equation (\ref{e4}) diverge,
if $\lambda>2$ both of them tend to zero. This case corresponds to 
$\rho \xrightarrow [t \to t_0]{} 0$, $|p| \xrightarrow [t \to t_0]{} \infty$ and we will consider it later.

Is a version of the ``Big Freeze'' with $t_0=\infty$ possible? It could be named the ``Little Freeze'' similarly to the situation with the 
``Big Rip'' and the ``Little Rip''. In this case instead of (\ref{e9}) we consider an asymptotic behaviour of the scale factor in the form 
$a(t)\xrightarrow [t \to \infty]{}a_0-Bt^{\lambda}$ with $\lambda<0$. According to (\ref{e2}) and (\ref{e3}) we have in this case 
$\rho(t)\propto t^{2\lambda-2}\xrightarrow [t \to \infty]{}0$ and $p(t)\propto t^{\lambda-2}\xrightarrow [t \to \infty]{}0$. This 
possibility will be considered later, too.

The only remaining singularity with infinite density is the well-known ``Big Crunch'' with $H \to -\infty$, which we do not consider 
here. 

\subsubsection{Finite terminal density}
Let us consider singularities with a nonsingular $\rho\xrightarrow [t \to t_0]{}\rho_0 \neq 0$. In this case all nontrivial solutions require
$p+\rho$ factor to diverge or vanish according to (\ref{e5}). In the first case $|p| \to \infty$, the second one corresponds to 
the crossing the line $\rho+p=0$. It corresponds to the equation of state of the cosmological constant, separating the phantom energy 
domain with an effective $w<-1$ from the domain of not so exotic matter $w>-1$. One can find in a literature both a statement that 
such a crossing is forbidden [\onlinecite{rr1}] and an example of a solution with such a crossing [\onlinecite{rr2}]. We will see that 
the possibility of such crossing depends on the parameters of the equation of state.

We start with considering solutions with finite $t_0$. Both cases could be described by a single power-law asymptote of the equation 
of state
\begin{equation}\label{e10}
\rho+p(\rho)\xrightarrow [\rho \to \rho_0]{}C(\rho-\rho_0)^\mu
\end{equation}
with $C=const$. 

At $\mu<0$ the modulus of the pressure tends to infinity, at $\mu>0$ the $\rho+p$ reaches zero.
The finiteness of $t_0$ is possible only at $\mu<1$. In this case we have $\rho(t)-\rho_0 \propto \Delta t^{1/(1-\mu)}, 
\rho(t)+p(t) \propto \Delta t^{\mu/(1-\mu)}$. The singularity with $\mu<0$ and $|p| \xrightarrow [\rho \to \rho_0]{} \infty$ is referred 
to as the type II or sudden singularity. The value of $H$ tends to finite $H_0$, so the scale factor linearly increases.

The achievement of $\rho+p=0$ condition in finite time is possible if $0<\mu<1$. Thus, the Universe can change the type of its equation of
state from phantom energy to a more ordinary one, but only for such kind of the asymptote of the equation of state.

At $\mu>1$ we obtain $t_0=\infty$, i.e. the asymptotic approximation of $\rho+p=0$ condition. The evolution of such a 
Universe at the terminal stage practically coincides with the evolution of the flat Universe with a 
cosmological constant
and without any other types of matter. There is no spacetime singularity in this case. Using the approximation (\ref{e10}) we obtain 
the asymptotes $\rho(t)-\rho_0 \propto t^{1/(1-\mu)}, \rho(t)+p(t) \propto t^{\mu/(1-\mu)}\to 0$ at $t \to \infty$. This solution 
can mimic the $\Lambda$CDM model.

\subsubsection{Zero terminal density}
This last possibility assumes $\rho_0=H_0=0$, which means that a scale factor tends to some extremum. But this does not means an 
asymptotic expansion or contraction of the Universe is impossible. One simple example is the case $a\propto t^\eta$, $0<\eta<1$ when the 
Universe keeps expanding, but $H$ decreases and tends to zero.

Let us consider the power-law asymptote of the equation of state
\begin{equation}\label{e11}
\rho+p\xrightarrow [\rho \to 0]{}-D\rho^\nu
\end{equation}
and substitute it into (\ref{e5}). The integral in (\ref{e5}) is finite at $\nu<1/2$, which yilds finite $t_0$. In this case 
$\rho \propto \Delta t^{2/(1-2\nu)}\xrightarrow [t \to t_0]{}0$, $H\propto \Delta t^{1/(1-2\nu)}\xrightarrow 
[t \to t_0]{}0$, $\rho+p\propto \Delta t^{2\nu/(1-2\nu)}$. If $0<\mu<1/2$, pressure tends to zero. This is a type IV singularity.
If $\lambda=1+1/(1-2\nu)$ is a noninteger number, the higher derivatives of $H\propto \Delta t^{\lambda-1}$ diverge. The condition 
$0<\mu<1/2$ means $\lambda>2$, so the first derivative of $H$ is finite, as well as both sides of the Friedmann equation (\ref{e4}). The
 value of $\lambda$ is the same as in (\ref{e9}). We can introduce the effective barotropic index $w=p/\rho\propto 
\Delta t^{(2\nu-2)/(1-2\nu)}\to \infty$. This singularity type was introduced in [\onlinecite{r10}].

If $\mu<0$ we have $|p|\xrightarrow [t \to t_0]{}\infty$. This is a new type of the future singularity, which we name ``Big Squeeze''. It 
combines certain properties of the sudden singularity and the type IV singularity. It corresponds to $1<\lambda<2$ in (\ref{e9}). The first 
derivative of $H$ and both sides of the Friedmann equation (\ref{e4}) diverge. The asymptotics near this singularity type are 
$\rho \propto \Delta t^{2/(1-2\nu)}\xrightarrow [t \to t_0]{}0$, $H\propto \Delta t^{1/(1-2\nu)}\xrightarrow 
[t \to t_0]{}0$, $|p|\propto \Delta t^{2\nu/(1-2\nu)}\xrightarrow [t \to t_0]{}\infty$, $a\xrightarrow [t \to t_0]{}a_0+
const \Delta t^{(2-2\nu)/(1-2\nu)}\to a_0$. It requires the equation of state (\ref{e11}) with negative $\nu$. The example is the
generalized Chaplygin gas which occurs in some cosmological theories.

At $1/2<\nu<1$ the integral in (\ref{e5}) diverges and $t_0=\infty$. In this case 
$\rho \propto t^{2/(1-2\nu)}\xrightarrow [t \to \infty]{}0$, $H\propto t^{1/(1-2\nu)}\xrightarrow 
[t \to \infty]{}0$, $\rho+p\propto t^{2\nu/(1-2\nu)}\xrightarrow [t \to \infty]{}0$, 
$a\xrightarrow [t \to \infty]{}a_0-Bt^{(2\nu-2)/(2\nu-1)}$. This is the mentioned above solution which could be named the ``Little 
Freeze''. In this case the effective barotropic index $w=p/\rho\propto t^{(2\nu-2)/(1-2\nu)}\to \infty$.
 
At $\nu=1/2$ we can take into account the possible logarithmic factor and consider the asymptotic equation of state
$\rho+p\xrightarrow [\rho \to 0]{}-D\rho^{1/2}(\ln \rho)^\beta$. At $\beta>1$ we deal with the unconventional type IV singularity
with $\ln\rho \propto \Delta t^{1/(1-\beta)}$, at $\beta<1$ we deal with the ``Little Freeze'' with $\ln\rho \propto t^{1/(1-\beta)}$.
At $\beta=1$ we consider the equation of state with the asymptotic $\rho+p\xrightarrow [\rho \to 0]{}-A\rho^{1/2}(\ln \rho)
(\ln \ln \rho)^\gamma$, etc. 

At $\nu>1$ we deal with the expanding Universe and $\ln a\propto t^{(2\nu-2)/(2\nu-1)}\xrightarrow [t \to \infty]{}\infty$ at $D>0$ in spite of 
$H\propto t^{1/(1-2\nu)}\xrightarrow [t \to \infty]{}0$. This is the new ``Little Freeze'' case. The higher derivatives of $H$ diverge. At
$\nu=1$ the Universe expands according to power law $a \propto t^{2/3D}$. The effective barotropic index $w=p/\rho \to -1$. 
Note that all version of the ``Little Freeze'' differ from 
the so-called pseudo-rip, which also corresponds to $t_0=\infty$ [\onlinecite{r9}].

\begin{table*}[tb]
\caption{Possible cosmological singularities except ``Big Bang'' and ``Big Crunch''}
\begin{tabular}{lccccccccc}
\hline
\hline
Type&Nickname&EoS&$\rho_0$&$|p_0|$&$p_0+\rho_0$&$a_0$&$\rho$&$p+\rho$&$a$\\
\hline
\multicolumn{10}{c}{$t \to t_0$, $\Delta t=t_0-t\to 0$} \\ \hline
I& ``Big Rip''&(\ref{e0}), $w<-1$&$\infty$&$\infty$&$-\infty$&$\infty$&
$\propto \Delta t^{-2}$&$\propto \Delta t^{-2}$&$a\propto \Delta t ^{-2/(3|1+w|)}$\\
I& ``Big Rip''&(\ref{e7}), $1/2<\alpha<1$&$\infty$&$\infty$&$-\infty$&$\infty$&
$\propto \Delta t^{2/(1-2\alpha)}$&$\propto \Delta t^{2\alpha/(1-2\alpha)}$&$\ln a\propto \Delta t^{2(1-\alpha)/(1-2\alpha)}$\\
III&``Big Freeze''&(\ref{e7}), $\alpha>1$&$\infty$&$\infty$&$-\infty$&$a_0$&
$\propto \Delta t^{2/(1-2\alpha)}$&$\propto \Delta t^{2\alpha/(1-2\alpha)}$&(\ref{e9}), $\lambda=(2\alpha-2)/(2\alpha-1)$\\
II& ``sudden''&(\ref{e10}), $\mu<0$&$\rho_0$&$\infty$&$-\infty$&$a_0$&
$\rho-\rho_0\propto \Delta t^{1/(1-\mu)}$&$\propto \Delta t^{\mu/(1-\mu)}$&$a\to a_0-H_0\Delta t$\\
IV&&(\ref{e11}), $0<\nu<1/2$&$0$&$0$&$0$&$a_0$&
$\propto \Delta t^{2/(1-2\nu)}$&$\propto \Delta t^{2\nu/(1-2\nu)}$&(\ref{e9}), $\lambda=(2-2\nu)/(1-2\nu)$\\
New&``Big Squeeze''&(\ref{e11}), $\nu<0$&$0$&$\infty$&$-\infty$&$a_0$&
$\propto \Delta t^{2/(1-2\nu)}$&$\propto \Delta t^{2\nu/(1-2\nu)}$&(\ref{e9}), $\lambda=(2-2\nu)/(1-2\nu)$\\
\hline
\multicolumn{10}{c}{$t \to \infty$} \\ \hline
& ``Little Rip''&(\ref{e7}), $0<\alpha<1/2$&$\infty$&$\infty$&$-\infty$&$\infty$&
$\propto t^{2/(1-2\alpha)}$&$\propto t^{2\alpha/(1-2\alpha)}$&$\ln a\propto t^{2(1-\alpha)/(1-2\alpha)}$\\
& ``Little Rip''&(\ref{e7}), $\alpha<0$&$\infty$&$\infty$&$0$&$\infty$&
$\propto t^{2/(1-2\alpha)}$&$\propto t^{2\alpha/(1-2\alpha)}$&$\ln a\propto t^{2(1-\alpha)/(1-2\alpha)}$\\
&``Little Freeze''&(\ref{e11}), $1/2<\nu<1$&$0$&$0$&$0$&$a_0$&
$\propto t^{2/(1-2\nu)}$&$\propto t^{2\nu/(1-2\nu)}$&$a\to a_0-Bt^{2(1-\nu)/(1-2\nu)}$\\
&``Little Freeze''&(\ref{e11}), $\nu>1$&$0$&$0$&$0$&$\infty$&
$\propto t^{2/(1-2\nu)}$&$\propto t^{2\nu/(1-2\nu)}$&$\ln a\propto t^{2(\nu-1)/(2\nu-1)}$\\ \hline \hline

\end{tabular}\label{t1}
\end{table*}

\subsection{Open and closed models} 
We went over the possible singularities for the case of the flat Universe and an arbitrary equation of state of its content. Let us study
the cases of open ($k=-1$) and closed ($k=1$) Universes. The second term in the right-hand side of (\ref{e2}) does not affects the 
properties of the singularities with $\rho, H \to \infty$  and $\rho\to\rho_0 \neq 0, H \to H_0 \neq 0$. The only exception is the ``Big Crunch'' singularity
with $a \to 0$ which we do not study in this paper. 

But we must revise a possibility of the existence and the properties of singularities
with $H \to 0$ or $\rho \to 0$. We use the asymptotic equation of state (\ref{e11}). All solutions with $a \to a_0$ are impossible because
the equation (\ref{e2}) cannot be satisfied. But the solutions with $a \to \infty$ can remain practically the same as in the flat 
case.  The matter with the asymptotic equation of state (\ref{e11}) with $\nu>1$ is a good example of this case. The term $ka^{-2}$ is much
less than the practically equal terms $H^2$ and $8\pi \rho /3$. At $\nu=1$ i.e. at $p\xrightarrow [\rho \to o]{}w\rho$ the term $ka^{-2}$
is much smaller than the other ones in (\ref{e2}) at $t \to \infty$ in the case $D>3/2$, i.e. $w<-5/2$. 

In the case $\nu<1$ we must compare the main terms in (\ref{e2}). The case $H^2~\rho\gg a^{-2}$ is impossible. The case $k=-1$, $H^2\approx
a^{-2}\gg \rho$ leads to $a=t$. This is the metric of the flat space-time and the coordinate transformation $r=t\sinh \xi, \tau=
t\cosh \xi$ turns it into the Minkowski metric. Naturally, this space-time is empty, $\rho=0$. However this is an unstable
solution. Considering some infinitesimal mass
density $\rho$ we get at $t \to \infty$ the asymptotic equation $d\rho/dt=3D\rho^{\nu}$ with the solution 
$\rho=\left(3D(1-\nu)\ln t+const\right)^{1/(1-\nu)}$. So $\rho$ diverges and the term with it becomes the main one in (\ref{e2}) at 
$t \to \infty$. The case $k=1$, $8\pi\rho/3\approx a^{-2}\gg H^2$ leads to $|da/dt|\ll 1$, $\rho \propto a^{-2}$. These conditions
exclude all known types of future singularities. Moreover, the last condition gives us $d\rho/dt=-2\rho H$. After substituting it into 
(\ref{e3}) we get $H(3D\rho^\nu -2\rho)=0$. This is possible if $H=0$ (the Einstein's static Universe) or $\rho=0$ (the empty Universe 
without DE) or $\nu=1, D=-2/3$. The last case means that $p=-\rho/3$. In all these cases we have no new type of singularity.

The last possibility is the case in which all terms in (\ref{e2}) are of the same order of magnitude. It gives us no solutions except 
the ```Big Crunch''. Thus, the equation of state (\ref{e11}) with $\nu<1$ could provide the type IV or the ``Big Squeeze'' singularities 
only for the flat model. There are no types of singularities specific for open or flat model.

\section{Conclusion}
We went over the possible types of future singularities for an arbitrary 
equation of state of the Universe with power-law asymptotes and found all the known types 
plus the new ``Big Squeeze'' and `` Little Freeze'' ones. We do not indicate all particular subtypes like ``the little sibling of the Big Rip 
singularity'' [\onlinecite{r11}], but find all the asymptotes for the scale factor, density and pressure in the vicinity of the main singularity types. 
We follow a unified approach. For the simplest flat model it reduces to the ordinary integral (\ref{e5}) for finite cosmological time 
of singularity or (\ref{e8}) for infinite one.

We tabulate all main cases of the cosmological singularities in Table \ref{t1}. The terminal values denoted $\rho_0$ and $a_0$ are 
finite and nonzero. EoS means the equation of state. Note that the ``Big Squeeze'' and the type IV cases are possible only for 
the flat Universe. The asymptote (\ref{e10}) of the equation of state at $0<\mu<1$ corresponds to changing the type of energy from 
phantom one to ordinary one or vice versa. At $\mu>1$ it provides an eternal near-$\Lambda$CDM expansion of the Universe. 

We consider naturally selected power-law types of equation of state for the all kinds of matter and DE in the Universe with the asymptotics 
(\ref{e0}), (\ref{e7}), (\ref{e10}) and (\ref{e11}) near the singularity. One can use the described approach in cases of more sophisticated 
equations of state with asymptotics like $\rho+p\xrightarrow [\rho \to \infty]{}-A\rho^\alpha\exp(u\rho^\eta)$, 
$\rho+p\xrightarrow [\rho \to \rho_0]{}C(\rho-\rho_0)^\mu\exp(u(\rho-\rho_0)^{-\eta})$ or 
$\rho+p\xrightarrow [\rho \to 0]{}-D\rho^\nu\exp(u\rho^{-\eta})$ with constant $u$ and $\eta$. Nevertheless, a simple
study of the integral (\ref{e5}) finiteness shows that they cannot provide an essential new type of cosmological singularity. The only 
things that change in these cases are the asymptotes of $a, \rho, p$, but the main properties of the singularity like the terminal density, pressure, and scale factor remain the same. So, 
the list of singularity types is exhaustive. 




\bibliography{big_rip}
\end{document}